\documentclass[aps,pra,reprint,superscriptaddress,showpacs]{revtex4-2}
\usepackage{bm}
\usepackage[dvipdfmx]{graphicx}
\usepackage{times}
\usepackage{color}
\usepackage{mathtools}
\usepackage{setspace}
\usepackage{here}

\newcommand{\sigda}{\sigma^{\dag}}
\newcommand{\akda}{a_k^{\dag}}
\newcommand{\ada}{a^{\dag}}
\newcommand{\bkda}{b_k^{\dag}}
\newcommand{\ain}{a_{\rm in}}
\newcommand{\aou}{a_{\rm out}}
\newcommand{\ainda}{a^{\dag}_{\rm in}}
\newcommand{\aouda}{a^{\dag}_{\rm out}}
\newcommand{\bin}{b_{\rm in}}
\newcommand{\bou}{b_{\rm out}}
\newcommand{\binda}{b^{\dag}_{\rm in}}
\newcommand{\bouda}{b^{\dag}_{\rm out}}
\newcommand{\cin}{c_{\rm in}}
\newcommand{\cou}{c_{\rm out}}
\newcommand{\cinda}{c^{\dag}_{\rm in}}
\newcommand{\couda}{c^{\dag}_{\rm out}}
\newcommand{\fin}{f_{\rm in}}
\newcommand{\fou}{f_{\rm out}}
\newcommand{\finco}{f^*_{\rm in}}
\newcommand{\fouco}{f^*_{\rm out}}
\newcommand{\Jp}{J_{+}}
\newcommand{\Jm}{J_{-}}
\newcommand{\Jpm}{J_{\pm}}
\newcommand{\Jz}{J_{z}}

\newcommand{\Xou}{X_{\rm out}}
\newcommand{\You}{Y_{\rm out}}

\newcommand{\gam}{\gamma}
\newcommand{\om}{\omega}
\newcommand{\oma}{\om_{\rm a}}
\newcommand{\omp}{\om_{\rm p}}

\newcommand{\Om}{\Omega}
\newcommand{\tp}{t_{\rm p}}
\newcommand{\Ein}{E_{\rm in}}
\newcommand{\cE}{{\cal E}}

\newcommand{\cEs}{{\cal E_{\rm S}}}
\newcommand{\Nin}{N_{\rm in}}
\newcommand{\Na}{N_{\rm a}}
\newcommand{\Nb}{N_{\rm b}}
\newcommand{\Nac}{N_{\rm ac}}
\newcommand{\Nbc}{N_{\rm bc}}
\newcommand{\Pa}{P_{\rm a}}
\newcommand{\Pb}{P_{\rm b}}
\newcommand{\Pac}{P_{\rm ac}}
\newcommand{\Pbc}{P_{\rm bc}}

\newcommand{\la}{\langle}
\newcommand{\ra}{\rangle}
\newcommand{\lp}{\left}
\newcommand{\rp}{\right}
\newcommand{\pri}{\prime}

\begin{document}

\title{Stimulated emission of superradiant atoms in waveguide QED}

\author{Rui Asaoka}
\email{rui.asaoka.st@hco.ntt.co.jp}
\affiliation{Computer and Data Science Laboratories, NTT Corporation, Musashino 180-8585, Japan}
\author{Julio Gea-Banacloche}
\email{jgeabana@uark.edu}
\affiliation{Department of Physics, University of Arkansas, Fayetteville, Arkansas 72701, USA}
\author{Yuuki Tokunaga}
\email{yuuki.tokunaga.bf@hco.ntt.co.jp}
\affiliation{Computer and Data Science Laboratories, NTT Corporation, Musashino 180-8585, Japan}
\author{Kazuki Koshino}
\email{kazuki.koshino@osamember.org}
\affiliation{College of Liberal Arts and Sciences, Tokyo Medical and Dental University, Ichikawa 272-0827, Japan}

\begin{abstract}
We investigate the stimulated emission of superradiant atoms coupled to a waveguide induced by a coherent-state photon pulse. We provide an analytical result when a short $\pi$ pulse is incident, which shows that the atoms emit photons coherently into the output pulse, which remains a coherent state in the short pulse limit. An incident pulse is amplified in phase-preserving manner, where noise is added almost entirely in the phase direction in phase space. This property improves the ratio of intensity signal to noise after the amplification for sufficiently short pulses. This is a unique feature different from general phase-preserving linear amplifiers, where the signal-to-noise ratio deteriorates in the amplification process. We also discuss the dependence of the photon-emission probability on pulse parameters, such as the pulse area and the duration.
\end{abstract}

\maketitle

Collective effects via the interaction between quantum emitters and reservoirs have attracted much attention as a consequence of quantum mechanics over the years and are now an indispensable part of modern technology. Especially, extensive efforts have been made to understand collective dynamics of atoms and {\it traveling} photons, which leads to novel opportunities for quantum networks and quantum information processing\,\cite{kimble,meter2016,simon}. The recent excitement of the field is due largely to the remarkable progress of the waveguide quantum electrodynamics (wQED) technology\,\cite{roy,chang,turschmann,sheremet}; it enables atoms couple {\it directly} to a waveguide, which differs strongly from the cavity-QED approaches of confining photons in all spatial directions.

Traveling photons and the radiation field from atoms coupling to a waveguide interfere with high coherence owing to the characteristic configuration of waveguides, namely the one-dimensional single-pass structure. The advantage has allowed experiments to observe fundamental optics of emitters\,\cite{astafiev,fujiwara}, which is also applied to optical devices\,\cite{hoi2011,loo,kannan}.
In the last several years, the controllable coupling of multiple atoms to a waveguide has also been realized with diverse platforms based on superconductors\,\cite{mirhosseini}, solid state defects\,\cite{sipahigil}, and cold atoms\,\cite{corozo2016,corozo2019}. In this novel experimental field, collective dynamics such as super- and sub-radiance has been observed\,\cite{goban,solano}.
On the other hand, in spite of the significant theoretical progress that began with Dicke's proposal of superradiance\,\cite{dicke}, we are still far from a complete understanding of the collective dynamics.

In this Letter, we investigate the stimulated emission of a cluster of excited atoms coupled to a waveguide.
Stimulated emission, widely known as the chief mechanism underlying lasers and masers\,\cite{gould,cummings,kogelnik,mollow}, is a paradigm of the collective effect which essentially reflects the indistinguishability of photons and multiphoton interference\,\cite{einstein,sun}.
In wQED systems, several studies have investigated the stimulated emission induced by a single-photon or Fock-state pulse\,\cite{rephaeli,valente,fischer}.
We consider the case that a coherent-state photon pulse is incident.
The stimulated emission induced by a coherent-state pulse is of fundamental interest from the perspective of how photons can be coherently added to a quantum field via its interaction with an atomic system\,\cite{hoi2012,zavatta}. More generally, it has the potential to be exploited not only for the coherent control of quantum emitters by pulses (or work extraction from a quantum system in a different context\,\cite{monsel}) but also vice-versa, i.e., the passive control of bosonic states through its interaction with quantum matter.

We provide an analytical result when a short $\pi$ pulse is incident, which shows that the atoms emit photons coherently into the waveguide. We find that the amplified pulse in this process remains a coherent state in the short pulse limit, not a non-classical state, such as the photon added coherent state (PACS)\,\cite{agarwal,francis}. 
We also characterize our system as an amplifier; the coherent-state pulse is amplified in phase-preserving manner\,\cite{cave}, where noise is added almost entirely in the phase direction in phase space. This amplification property lets the ratio of intensity signal to noise improve after the amplification for sufficiently short pulses. This is a unique feature different from general phase-preserving linear amplifiers, where the signal-to-noise ratio in the output deteriorates compared to that of the input\,\cite{cave}.

\begin{figure}
\includegraphics[clip,width=8cm]{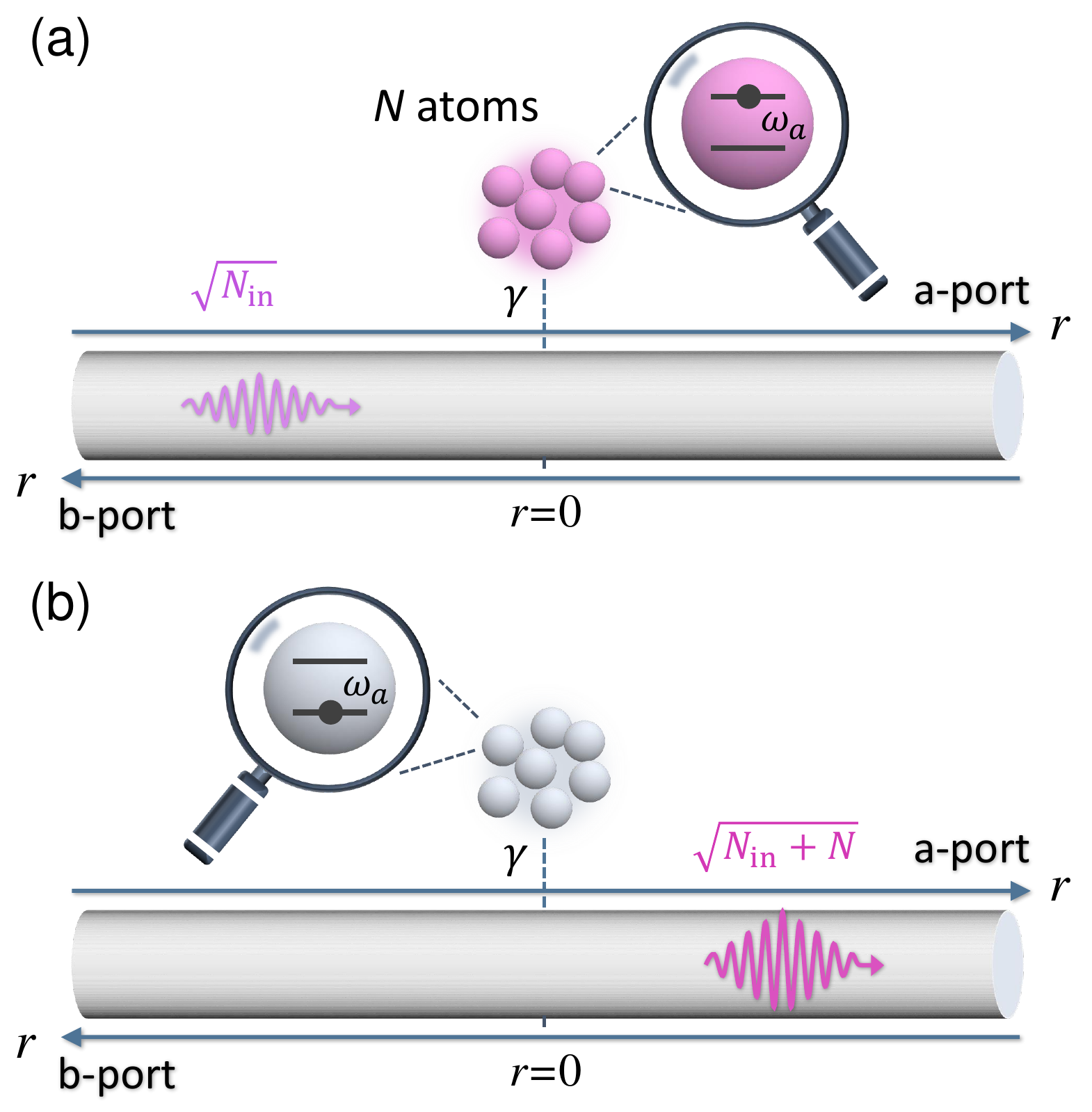}
\caption{Schematic of an infinite waveguide coupled to $N$ two-level atoms. Photons propagate forward ($a$-port) or backward ($b$-port) along the waveguide. (a) Initial state: all atoms are excited and a coherent-state pulse $|\sqrt{\Nin}\ra$ propagating forward is input. (b) Final state: under some conditions (see text), all atoms coherently emit photons into the incident pulse, resulting in a grown coherent state, $|\sqrt{\Nin + N}\ra$.}
\label{fig:wagu}
\end{figure}

Here we consider an infinitely long waveguide that couples with $N$ identical two-level atoms (transition frequency $\oma$, radiative decay rate $\gam$ into the waveguide) at the same position (${r=0}$), as illustrated in Fig.\,\ref{fig:wagu}(a). In such a situation, the atoms interact with the input pulse in a collective fashion; that is, superradiance is expected to occur.
%
%
%
Denoting the lowering operator of $j$th atom by $\sigma_j$ and the annihilation operator of a waveguide photon 
propagating forward (backward) with wavenumber $k\,(>0)$ by $a_k$ ($b_k$), the Hamiltonian of the overall system is given by
%
\begin{align}
H = \oma \Jz
&+ \int^{\infty}_{-\infty} dk \lp[k\akda a_k + i\sqrt{\frac{\gam}{4\pi}} 
(\Jp a_k - \akda \Jm) \rp]
\nonumber
\\
&+ \int^{\infty}_{-\infty} dk \lp[k\bkda b_k + i\sqrt{\frac{\gam}{4\pi}} 
(\Jp b_k - \bkda \Jm) \rp]   ,
\label{ham}
\end{align}
where $\Jpm$ and $\Jz$ are the collective spin operators of the $N$ atoms defined as ${\Jm=\sum_j\sigma_j}$, ${\Jp=\Jm^{\dag}}$, and ${\Jz=\sum_j\sigda_j\sigma_j - N/2}$.
Note the following two points: 
(i)\,We assume a linear dispersion $\om_k=vk$ (group velocity $v$)
for the waveguide photons and set $v=1$. 
(ii)\,We extended the lower limit of the $k$-integral in Eq.\,(\ref{ham}) from $0$ to $-\infty$. 
This introduces photons with negative energies, 
but they are irrelevant in this model due to large detuning from the atoms.

Initially (${t=0}$), we assume that all the atoms are in the excited state, and a coherent-state pulse propagating forward is incident to the excited atoms (Fig.\,\ref{fig:wagu}(a)). We denote the amplitude of the pulse at the atomic position by $\Ein(t)$. The initial state of the total system is then written as
%
\begin{align}
|\psi(0)\ra &= W\exp\lp(\int_{-\infty}^{0} dr
\Ein(-r)\tilde{a}_r^{\dag}\rp)\prod_{j=1}^{N}\sigda_j|{\rm vac}\ra   ,
\label{1ini}
\end{align}
where $|{\rm vac}\ra$ is the ground state of the overall system, and the normalization constant is given by $W=\exp(-\frac{1}{2}\int^0_{-\infty} dr|E_{\rm in}(-r)|^2)$. The real-space representation $\widetilde{a}_r$ of the waveguide operator is defined as the Fourier transform of $a_k$ by ${\tilde{a}_{r}(t) \equiv \frac{1}{\sqrt{2\pi}}\int dk a_k e^{ikr}}$. In particular, the input and output field operators are defined by $\ain(t) = \tilde{a}_{-0}(t)$ and $\aou(t) = \tilde{a}_{+0}(t)$.
$\widetilde{b}_r(t)$, $\bin(t)$, and $\bou(t)$ are defined similarly.
The Heisenberg equations of the collective spin operators are given by
%
\begin{align}
\frac{d}{dt}\Jm &= -\lp(i\oma + \gam\rp)\Jm + \gam\Jm\Jz
\nonumber
\\
& \hspace{0.4cm} - \sqrt{2\gam}\Jz[\ain(t) + \bin(t)]   ,
\label{eq:dsdt}
\\
\nonumber
\frac{d}{dt}\Jz & = 
- \gam\Jp\Jm
+ \sqrt{\frac{\gam}{2}}[\ainda(t) + \binda(t)]\Jm\\
& \hspace{0.4cm} + \sqrt{\frac{\gam}{2}}\Jp[\ain(t) + \bin(t)]   ,
\label{eq:dzdt}
\end{align}
and the input-output relations are given by
%
\begin{align}
\aou(t) &= \ain(t) - \sqrt{\frac{\gam}{2}}\Jm   ,
\label{aout}
\\
\bou(t) &= \bin(t) - \sqrt{\frac{\gam}{2}}\Jm   .
\label{bout}
\end{align}

We denote the expectation value of an operator $\hat{o}(t)$ by $\la \hat{o}(t)\ra=\la\psi(0)| \hat{o}(t)|\psi(0)\ra$.
From Eqs.\,(\ref{1ini})-(\ref{eq:dzdt}), we obtain
%
\begin{align}
\frac{d}{dt}\la\Jm\ra &= 
- \lp(i\oma + \gam\rp)\la\Jm\ra + \gam\la\Jm\Jz\ra
\nonumber
\\
& \hspace{0.4cm} - \sqrt{2\gam}\Ein(t)\la\Jz\ra   ,
\label{eq:dsdt2}
\\
\frac{d}{dt}\la\Jz\ra & = 
- \gam\la\Jp\Jm\ra
+ \sqrt{\frac{\gam}{2}}\Ein^*(t)\la\Jm\ra
\nonumber
\\
& \hspace{0.4cm} + \sqrt{\frac{\gam}{2}}\Ein(t)\la\Jp\ra   ,
\label{eq:dzdt2}
\end{align}
with the initial conditions $\la\Jm\ra = 0$ and $\la\Jz\ra = N/2$. In deriving above equations, we used ${\ain(t)|\psi(0)\ra = \Ein(t)|\psi(0)\ra}$ and ${\bin(t)|\psi(0)\ra = 0}$.

\begin{figure}
\includegraphics[clip,width=8.5cm]{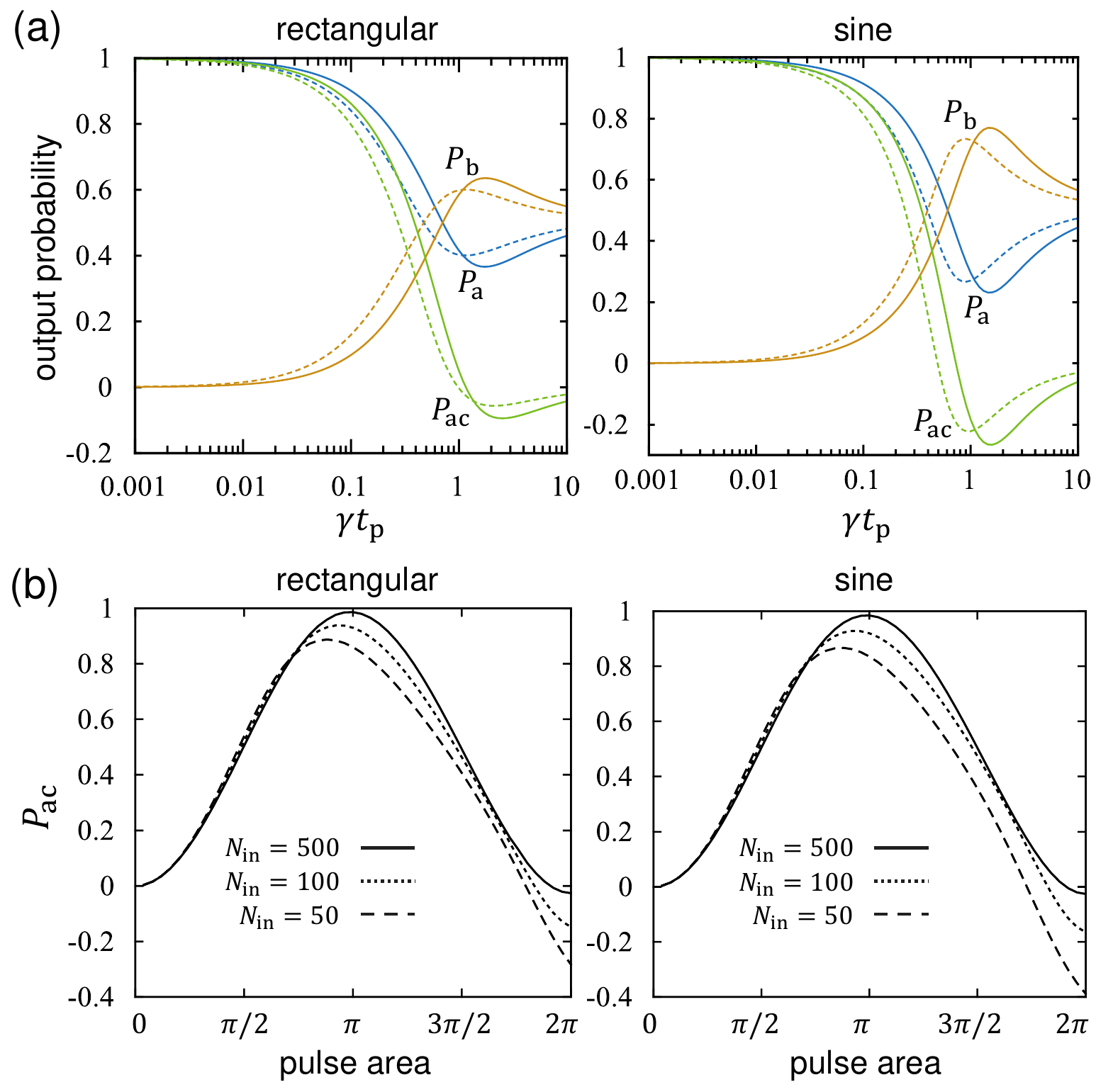}
\caption{Photon emission probabilities for rectangular and sine input pulses. (a) Probabilities of photons being output into the a(b)-port $\Pa$($\Pb$) and the coherent-emission probability $\Pac$ as a function of the $\pi$-pulse length $\tp$. Solid and dashed curves are the probabilities for $N=10$ and $N=20$, respectively. (b) Coherent-emission probability as a function of the area of an input pulse $A$ for the atom number of ${N=10}$.}
\label{fig:pro}
\end{figure}

The mean number of input photons is given by $\Nin = \int_0^{\infty} dt \la\ainda\ain\ra$.
The mean number of output photons in the forward direction and its coherent component are respectively given by $\Na = \int_0^{\infty} dt \la\aouda\aou\ra$ and $\Nac = \int_0^{\infty} dt |\la\aou\ra|^2$. $\Nb$ and $\Nbc$ are defined similarly.
The system dynamics obeying Eqs.\,(\ref{eq:dsdt2}) and (\ref{eq:dzdt2}) conserves the total photon number as ${\Na + \Nb = \Nin + N}$\,\cite{sup1}.
From these quantities, we define the forward/backward photon emission probabilities:
\begin{align}
\Pa &= (\Na-\Nin)/N   ,
\label{pa}
\\
\Pac &= (\Nac-\Nin)/N   ,
\label{pac}
\\
\Pb &= \Nb/N.
\label{pb}
\end{align}
Note that $\Pa$ and $\Pac$ may take negative values when the output photon number in the forward direction is decreased 
in comparison with the input photon number, as we will see in Fig.\,\ref{fig:pro}.

First, we consider the case in which a rectangular $\pi$ pulse is input.
$\Ein (t)$ is given by
\begin{align}
\Ein(t) = 
\begin{cases}
\frac{\Omega}{\sqrt{2\gam}} e^{-i(\oma t-\theta)} & (0 \leq t \leq \tp)
\\
0 & ({\rm otherwise})
\end{cases},
\label{recpul}
\end{align}
where $\Omega$ is the Rabi frequency and $\theta$ is the phase of input pulse.
From the condition that the pulse area (defined by $A=\sqrt{2\gam}\lp|\int dt\Ein(t)e^{i\oma t}\rp|$) is equal to $\pi$, $\Om$ and $\tp$ are related by $\Om \tp=\pi$.
The mean number of input photons is given by 
\begin{align}
\Nin=\pi\Om/(2\gam)=\pi^2/(2\gam \tp).
\label{Nin}
\end{align}
In particular, we focus on the case of a short $\pi$ pulse
satisfying ${\tp \ll \gam^{-1}}$ (namely, ${\gam \ll \Om}$). In that case, we can neglect damping of atoms during the $\pi$-pulse duration. 
Then, Eqs.\,(\ref{eq:dsdt2}) and (\ref{eq:dzdt2}) reduce to 
$\frac{d}{dt}\la\Jm\ra = -i\oma\la\Jm\ra-\Om e^{-i(\om_a t-\theta)} \la\Jz\ra$ and
$\frac{d}{dt}\la\Jz\ra = \frac{\Om}{2}[e^{i(\om_a t-\theta)}\la\Jm\ra + \mathrm{c.c.}]$, respectively. 
With the initial conditions of $\la\Jz(0)\ra = N/2$ and $\la\Jm(0)\ra = 0$, these differential equations are solved as
%
\begin{align}
\la\Jm\ra &=
\begin{cases}
-\frac{N}{2} \sin(\Om t)e^{-i(\oma t-\theta)} & 
(0 \leq t \leq \pi/\Om)
\\
0 & (\rm{otherwise})
\end{cases}   ,
\label{sol:g1}
\\
\la\Jz\ra &=
\begin{cases}
\frac{N}{2}\cos(\Om t) & (0 \leq t \leq \pi/\Om)
\\
0 & (\rm{otherwise})
\label{sol:g2}
\end{cases}.
\end{align}
From the above solutions, we can evaluate the mean number of output photons, $\Na$, $\Nac$, $\Nb$, and $\Nbc$. 
For example, the amplitude of the output photons in the forward direction is derivable from the input-output relation (\ref{aout}) as
%
\begin{align}
&\la \aou(t) \ra =
\nonumber
\\
&\begin{cases}
\lp[\frac{\Om}{\sqrt{\gam}} + N\sqrt{\frac{\gam}{8}}\sin(\Om t)\rp] 
e^{-i(\oma t-\theta)} & (0 \leq t \leq \pi/\Om)
\\
0 & ({\rm otherwise})
\end{cases}.
\label{aout_ana}
\end{align}
By integrating the squared amplitude, $\Nac$ is given by
\begin{align}
\Nac &= \Nin + N + {\cal O}(\gam/\Om), 
\label{num_aout}
\end{align}
where ${\cal O}(\gam/\Om)$ represents a small quantity of the order of $\gam/\Om$, which vanishes in the short $\pi$-pulse limit.
Following the similar arguments, we confirm that
\begin{align}
\Na &= \Nac = \Nin + N + {\cal O}(\gam/\Om), 
\\
\Nb &= \Nbc = {\cal O}(\gam/\Om).  
\end{align}
Namely, ${\Pa=\Pac=1}$ and ${\Pb=\Pbc=0}$ in the short pulse limit. The fact that ${\Pa=1}$ and ${\Pb=0}$ implies that all $N$ atoms emit photons to the forward direction, stimulated by the input $\pi$ pulse. 
Since the number of photons emitted by the atoms has no fluctuations, one may expect that the output pulse is in a $N$-PACS\,\cite{agarwal,francis}; it is known that the single-PACS ($N=1$) is generated via the stimulated emission in an optical parametric amplifier (OPA)\,\cite{zavatta}. $N$-PACS is defined by $(\ada)^n |\alpha\ra$ ($|\alpha\ra$ denotes a coherent state and normalization constant is omitted), which exhibits several non-classical characteristics\,\cite{agarwal,francis}.
However, the fact that $\Na=\Nac$ implies that the output pulse is in a coherent state with an increased mean photon number, $|\sqrt{\Nin+N}\ra$ [Fig.\,\ref{fig:wagu}(b)].
This is proved by the fact that the number of coherent photons, $|\la a \ra|^2$, is substantially different between the amplified coherent state and the $N$-PACS, as shown in the Supplemental Materials\,\cite{sup2}.
Thus, the {\it pure} stimulated emission in our system is a quite different process from the photon addition through an OPA, where an incident coherent pulse goes through a somewhat complicated process with significantly weak crystal-light interaction and the projection by the detection of the idler photon.

Stimulated emission by propagating field is highly sensitive to the pulse area of the input field. For example, if the input pulse is a short $2\pi$-pulse, it re-excites all $N$ atoms and transmits without the growth of the photon number. 
The radiative decay of the atoms (superradiance) occur afterwards, emitting incoherent photons in both directions with equal probability. In that case, one has ${\Na=\Nin+N/2}$, ${\Nac=\Nin}$, ${\Nb=N/2}$, and ${\Nbc=0}$.

\begin{figure}
\includegraphics[clip,width=7cm]{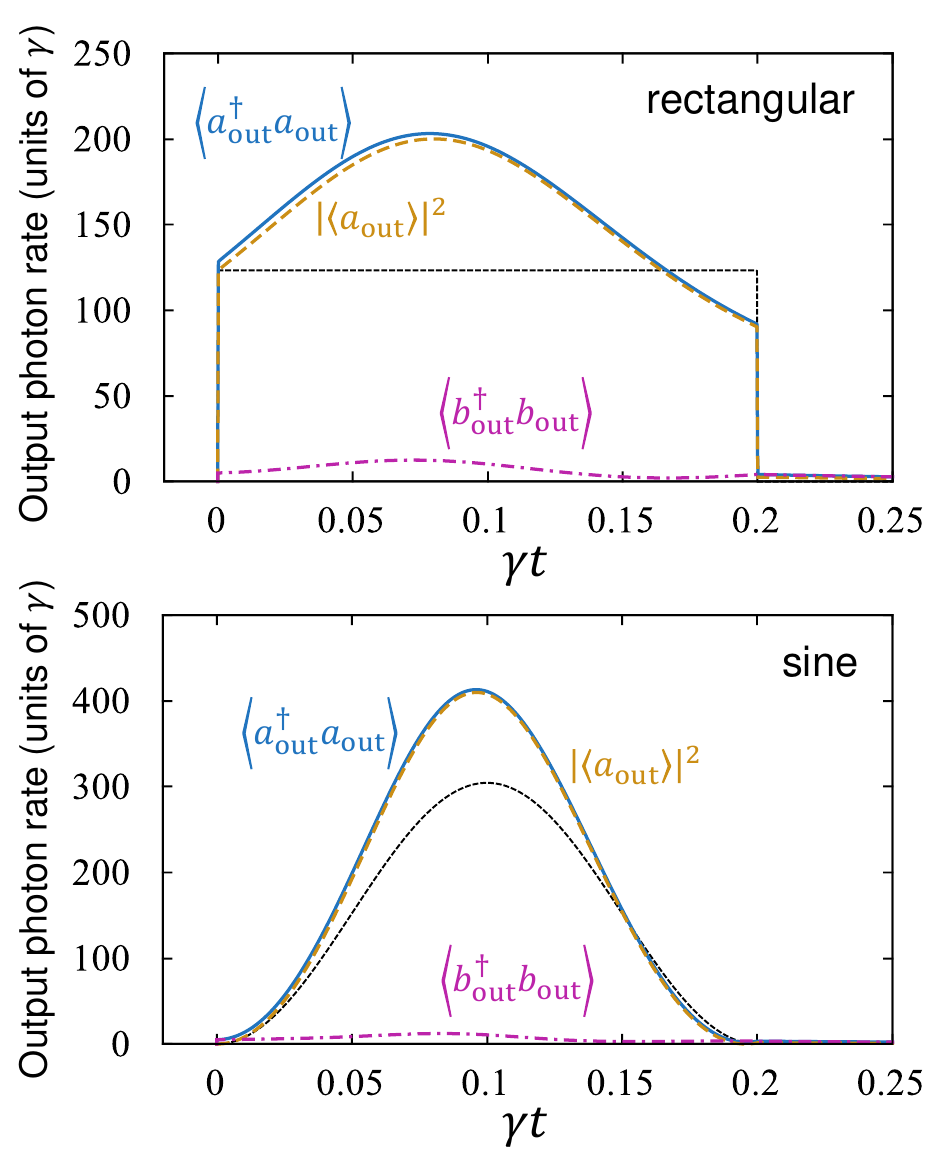}
\caption{Shape of the input and output pulses: 
$\la\ainda\ain\ra(=|\la\ain\ra|^2)$ (dotted), $\la\aouda\aou\ra$ (solid), $|\la\aou\ra|^2$ (dashed), and $\la\bouda\bou\ra$ (dashed-dotted).
Note that the solid and dashed curves are almost overlapping in the sine-pulse case. 
The system parameters are set to ${\tp = 0.2\gam^{-1}}$, and ${N = 10}$.
The photon numbers included in the pulses are $(\Nin,\Na,\Nac,\Nb)=(24.66,32.77,31.95,1.87)$ (rectangular pulse) and $(30.44,38.72,37.69,1.70)$ (sine pulse).}
\label{fig:shape}
\end{figure}

Next, we rigorously simulate the dynamics of the atoms according to the Hamiltonian (\ref{ham})\,\cite{sup3} and investigate the stimulated emission more quantitatively beyond the short pulse limit.
Note that the Hamiltonian of Eq.\,(\ref{ham}) conserves the total angular momentum ${\bf J}^2=J_x^2+J_y^2+J_z^2$ of the collective atom operator,
and the state vector of atoms evolve within the Hilbert space of a fixed azimuthal quantum number, $l=N/2$\,\cite{sup3}.
Besides a rectangular pulse, we also consider a sine pulse for comparison, which is given by 
\begin{align}
\Ein(t) &=
\begin{cases}
\cEs\sin\lp(\frac{\pi t}{\tp}\rp) e^{-i(\oma t-\theta)} & (0 \leq t \leq \tp)
\\
0 & ({\rm otherwise})
\end{cases},
\label{sinpul}
\end{align}
where $\cEs=\frac{\pi^2}{2\tp\sqrt{2\gam}}$ for the $\pi$-pulse case. The mean input photon number is given by ${\Nin=\pi^4/(16\gam\tp)}$. Figure\,\ref{fig:shape} plots the shapes of the output pulses in $a$-port and $b$-port. We confirm that, regardless of the input pulse shape,  
$\la\aouda\aou\ra$ and $|\la\aou\ra|^2$ are mostly overlapping, and they are much larger than $\la\bouda\bou\ra$. 
This agrees with our previous observation that most atoms emit photons coherently into the forward direction when the input pulse length is much shorter than the lifetime of atoms ($\tp \ll \gam^{-1}$).
For a rectangular input pulse, reflecting the discontinuity of the pulse at $t=0$ and $\tp$, the shapes of the input and output pulses are substantially deformed. In contrast, for a sine input pulse, the input pulse is amplified by the excited atoms without substantial deformation of the pulse shape.
%
%
%

Figure\,\ref{fig:pro}(a) plots the photon emission probabilities as a functions of the $\pi$-pulse length $\tp$.
In the short $\pi$-pulse region (${\tp\lesssim 0.1\gam^{-1}}$), the atoms emit photons mostly in the $a$-port as a coherent component (${\Pa=\Pac\approx 1}$) in agreement with our preceding observation.
By comparing results for $N=10$ and $20$ (solid and dotted lines in Fig.\,\ref{fig:pro}(a)), we can also confirm that $\Pa$ and $\Pac$ are decreased for larger atom number $N$. This is explained as follows. For larger $N$, the radiative decay of atoms becomes faster due to the superradiance. Therefore, a shorter $\pi$ pulse is required to induce complete stimulated emission.

As the pulse length increases, the output probability into $b$-port increases due to the coherent reflection of input photons: In the long $\pi$-pulse region ($\tp \gg \gam^{-1}$), the radiative decay of the atoms is almost unaffected by the input pulse.
Near the initial moment ($t \lesssim \gam^{-1}$), the atoms quickly emit incoherent photons into the $a$- and $b$-ports with equal probabilities (superradiance). Most of the input photons arrive afterwards and are reflected coherently by the atoms, which are now completely deexcited\,\cite{astafiev}.
Therefore, ${\Na=N/2}$, ${\Nac=0}$, ${\Nb=N/2+\Nin}$, and ${\Nbc=\Nin}$. 
For the rectangular pulse case where ${\Nin=\pi^2/(2\gam \tp)}$, we have ${\Pa = 1/2-\pi^2/2N\gam\tp}$, ${\Pb = 1/2+\pi^2/2N\gam\tp}$, and ${\Pac = -\pi^2/2N\gam\tp}$. We have confirmed that these are in good agreement with Fig.\,\ref{fig:pro}(a).

In Fig.\,\ref{fig:pro}(b), the coherent-emission probability $\Pac$ is shown as a function of the input pulse area $A$, fixing the mean photon number of the input pulse $\Nin$. For a pulse including sufficient photons, $\Pac$ takes a maximum value close to unity around ${A = \pi}$.
$\Pac$ decreases for smaller $\Nin$. This is because, for a fixed pulse area, the pulse length becomes longer for smaller $\Nin$ ($\tp=A/(2\gam\Nin)$ in the rectangular case), which causes the increase of incoherent photons due to spontaneous emission during the pulse duration.
We observe in Fig.\,\ref{fig:pro}(b) that there is little difference between the rectangular and sine pulses. Thus, the difference of the pulse shape is not essential for the stimulated emission.
%
%
%

Finally, we observe the quantum state of the output pulse and characterize the stimulated emission as an amplification process. We define the input and output pulse mode operators by
%
\begin{align}
\cin &= \int^\infty_0 dt\finco(t)\ain(t)   ,
\label{cin}
\\
\cou &= \int^\infty_0 dt\fouco(t)\aou(t)   ,
\label{cout}
\end{align}
where $\fin(t)$ ($\fou(t)$) is the mode function of the input (output) pulse. In order that $c_j$ (${j={\rm in, out}}$) satisfies the commutation relation ${[c_j, c^{\dag}_j] = 1}$ for a discrete boson, $f_j(t)$ is normalized as ${\int^\infty_0 dt |f_j(t)|^2 = 1}$.
We choose the output mode function to be identical with the input one. 
For a rectangular pulse case for example, they are given by
%
\begin{align}
\fin(t)=\fou(t)=
\begin{cases}
\sqrt{1/\tp}\,e^{-i\oma t} & (0<t<\tp)
\\
0 & ({\rm otherwise})
\end{cases}   .
\label{fdef}
\end{align}
Regarding the input pulse, we can readily check the following coherent-state property, ${\la{\cinda}^\alpha\cin^\beta\ra = {\la\cin\ra^*}^\alpha\la\cin\ra^\beta}$, where ${\la\cin\ra = \sqrt{\Nin}e^{i\theta}}$.

\begin{figure*}[t]
\includegraphics[clip,width=17.5cm]{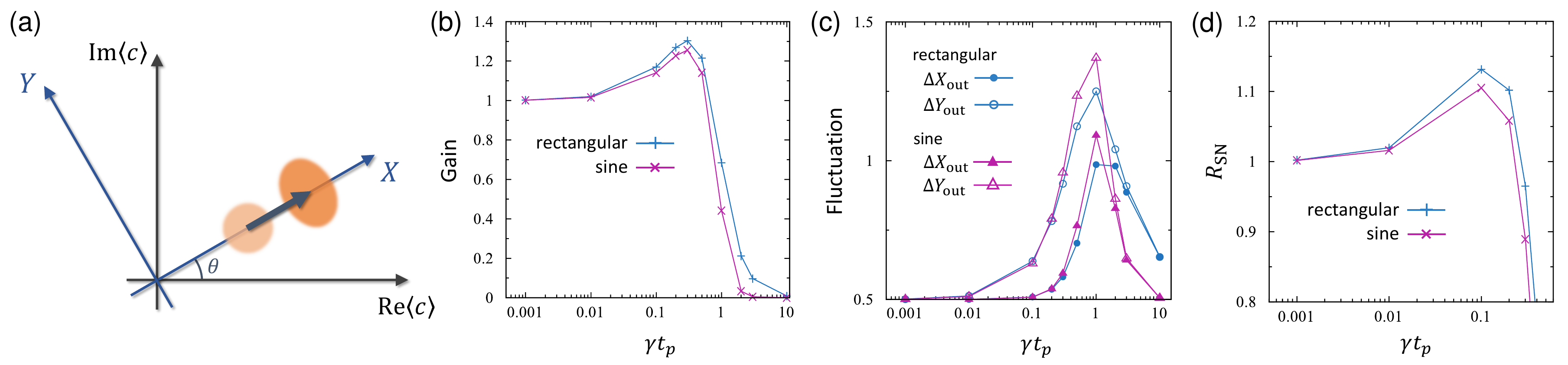}
\caption{(a) Definition of the quadrature axes and illustration of the amplification process of a coherent-state pulse in phase space. (b) Signal gain defined as ${G \equiv \la\cou\ra^2/\la\cin\ra^2}$ as a function of the pulse length of rectangular and sine $\pi$ pulses. (c) Quadrature fluctuations in the output pulse for the same input as (b). (d) Rate of change in the ratio of the intensity signal to noise before and after the amplification process, defined by $R_{\rm SN}\equiv[\la\cou\ra^2/(\Delta\cou)^2]/[\la\cin\ra^2/(\Delta\cin)^2]=G/[4(\Delta\Xou)^2]$, for the same input parameters as in (b) and (c). In (b)-(d), we set the atom number ${N=10}$, and the results are rotationally symmetric in phase space, i.e., independent of the angle $\theta$ in (a).}
\label{fig:flu}
\end{figure*}

We define the quadrature components by $\Xou = (\cou e^{-i\theta} + \couda e^{i\theta})/2$ and $\You = (\cou e^{-i\theta} - \couda e^{i\theta})/2i$ (Fig.\,\ref{fig:flu}(a)). Their fluctuations are given by $(\Delta\Xou)^2 = (1 + 2\la\couda,\cou\ra + \la\cou,\cou\ra e^{-2i\theta} + \la\couda,\couda\ra e^{2i\theta})/4$ and $(\Delta\You)^2 = (1 + 2\la\couda,\cou\ra - \la\cou,\cou\ra e^{-2i\theta} - \la\couda,\couda\ra e^{2i\theta})/4$, where $\la \hat{A},\hat{B}\ra \equiv \la \hat{A}\hat{B}\ra - \la \hat{A}\ra\la \hat{B}\ra$. Details on evaluation of the quadrature fluctuations are presented in the Supplemental Materials\,\cite{sup4}. We regard complex amplitudes, $\la\cin\ra$ and $\la\cou\ra$, as the signal and define ${G \equiv \la\cou\ra^2/\la\cin\ra^2}$ as the signal gain. We can readily check that $\la\cin\ra e^{-i\theta}$ and $\la\cou\ra e^{-i\theta}$ are both real quantities, and consequently $G$ reduces to a real quantity in our definition. At the same time, we can confirm that the signal gain $G$ is insensitive to the phase $\theta$ of the input pulse, in contrast with the phase-sensitive amplification such as the parametric amplification.
%
%

Figure\,\ref{fig:flu}(b) shows the dependence of the signal gain on the input $\pi$-pulse length $\tp$. As we observed earlier, in the short pulse limit, all atoms coherently emit photons into the incident pulse, i.e., ${\la\cou\ra=\sqrt{\Nin+N}\,e^{i\theta}}$ when ${\la\cin\ra=\sqrt{\Nin}\,e^{i\theta}}$. 
However, as the input photon number $\Nin$ increases for a shorter $\pi$ pulse (for example, see Eq.\,(\ref{Nin})), the gain defined here decreases and approaches unity as ${G=1+2N\gam\tp/\pi^2}$.
Regardless of the pulse shape, the gain takes the maximum value around ${\tp\approx0.3/\gam}$ and drops sharply for ${\tp\gtrsim 1/\gam}$.
The loss of signal magnitude is due to the coherent reflection of the input pulse\,\cite{astafiev}.

Figure\,\ref{fig:flu}(c) shows the dependence of the quadrature fluctuations on the input $\pi$-pulse length $\tp$. A remarkable point here is that, in the short pulse region satisfying ${\gam\tp\lesssim 0.1}$, almost no noise is added to the $X$ quadrature, although the signal is amplified (${G > 1}$).
Within the linear amplification framework, it has been known that the amplification must add noise as $(\Delta \hat{v}_{\rm out})^2 = G_v(\Delta \hat{v}_{\rm in})^2 + (\Delta\hat{F})^2$ ($G_v$ is the gain with regards to a mode $\hat{v}$ and $\Delta\hat{F}$ is noise caused by internal modes of an amplifier)\,\cite{cave}. Therefore, signal-to-noise ratio, which we define here by $\la\hat{v}\ra^2/(\Delta\hat{v})^2$, deteriorates in an amplification process as $\la\hat{v}_{\rm out}\ra^2/(\Delta\hat{v}_{\rm out})^2 = G_v\la\hat{v}_{\rm in}\ra^2/{[G_v(\Delta\hat{v}_{\rm in})^2+(\Delta\hat{F})^2]} < \la\hat{v}_{\rm in}\ra^2/(\Delta\hat{v}_{\rm in})^2$, where we assumed $\la\hat{F}\ra = 0$. In contrast, in our study, the signal-to-noise ratio of the output, $\la\cou\ra^2/(\Delta\cou)^2$, is improved compared to that of the input in the short-pulse region, as shown in Fig.\,\ref{fig:flu}(d)\,\cite{exp}.
In the $Y$ quadrature (phase direction), on the other hand, the stimulated emission adds substantial noise as illustrated in Fig\,\ref{fig:flu}(a).
These properties would enable the control of excitation number of bosonic states in quantum information processing, in the phase preserving manner and with minimal change in the photon-number fluctuation. It would be advantageous in some bosonic-encoding schemes, such as logical qubits with the rotation-symmetric bosonic states or the superposition of coherent states\,\cite{grimsmo} with different phases\,\cite{leghtas,grimm}.

In summary, we have investigated the stimulated emission of superradiant atoms coupled to a waveguide induced by a coherent-state pulse. We have shown that a short $\pi$ pulse induces the coherent photon emission of the atoms into the output pulse, which remains a coherent state in the short pulse limit. The pulse amplification in our system is phase-preserving, where noise is added almost entirely in the phase direction in phase space. This property improves the signal-to-noise ratio after the amplification for sufficiently short pulses.
Our study would be helpful for building the quantum network and provides insights into the fundamental physics based on the interaction between a quantum field and an atomic system, which may lead to passive control of bosonic states through its interaction with quantum matter in quantum processing.

\begin{acknowledgments}
The authors are grateful to S. Kono and Y. Kubo for fruitful discussions.
This work was supported by JST CREST (Grant No. JPMJCR1775), JST ERATO (Grant No. JPMJER1601), and JSPS KAKENHI (Grant No. 19K03684).
\end{acknowledgments}

\newpage
\widetext
\begin{center}
\textbf{\large Supplementary materials for ``Stimulated emission of superradiant atoms in waveguide QED''}
\end{center}
\setcounter{section}{0}
\setcounter{equation}{0}
\setcounter{figure}{0}
\setcounter{table}{0}
\setcounter{page}{1}
\makeatletter
\renewcommand{\theequation}{S\arabic{equation}}
\renewcommand{\thefigure}{S\arabic{figure}}
\renewcommand{\bibnumfmt}[1]{[S#1]}
\renewcommand{\citenumfont}[1]{S#1}

\section{Conservation of total photon number}
\label{sup:connum}
We can verify the conservation of the total photon number before and after the atom-pulse interaction as follows.
%
\begin{align}
\Na + \Nb
&=\int_0^{\infty}dt \la\aouda(t)\aou(t)\ra + \int_0^{\infty}dt\la\bouda(t)\bou(t)\ra
\nonumber
\\
&= \int_0^{\infty}dt \la\ainda(t)\ain(t)\ra + \int_0^{\infty}dt\la\binda(t)\bin(t)\ra
+ \gam\int_0^{\infty}dt\la\Jp\Jm\ra
- \sqrt{\frac{\gam}{2}}\int_0^{\infty}dt\lp[\Ein^*(t)\la\Jm\ra + \Ein(t)\la\Jp\ra\rp]
\nonumber
\\
&=\Nin - \int_0^{\infty}dt\lp[\frac{d}{dt}\la\Jz\ra\rp]
\nonumber
\\
&=\Nin + \la\Jz(0)\ra - \la\Jz(\infty)\ra
\nonumber
\\
&= \Nin + N   .
\label{conse}
\end{align}
%
\section{Coherence of photon added state}
\label{sup:pacs}
Here we show the discrepancy between the coherent photon numbers $|\la a\ra|^2$ in the coherent state with increased photon number $|\sqrt{\Nin+N}\ra$ and in the $N$-PACS defined as ${\lp|\sqrt{\Nin},N\rp\ra \equiv \frac{\lp(\ada\rp)^N}{\sqrt{\lp\la\sqrt{\Nin}\rp|a^N(\ada)^N\lp|\sqrt{\Nin}\rp\ra}}\lp|\sqrt{\Nin}\rp\ra}$. The coherence of $N$-PACS is given by
%
\begin{align}
\lp\la\sqrt{\Nin},N\rp|a\lp|\sqrt{\Nin},N\rp\ra =
\frac{\lp\la\sqrt{\Nin}\rp|a^{N+1}\lp(\ada\rp)^N\lp|\sqrt{\Nin}\rp\ra}{\lp\la\sqrt{\Nin}\rp|a^N(\ada)^N\lp|\sqrt{\Nin}\rp\ra}.
\label{cpacs}
\end{align}
The numerator in Eq.\,(\ref{cpacs}) is calculated as
%
\begin{align}
\nonumber
\lp\la\sqrt{\Nin}\rp|a^{N+1}\lp(\ada\rp)^N\lp|\sqrt{\Nin}\rp\ra
&= \lp\la 0\lp|D^{\dag}\lp(\sqrt{\Nin}\rp)a^{N+1}\lp(\ada\rp)^ND\lp(\sqrt{\Nin}\rp)\rp|0\rp\ra \\
\nonumber
&= \lp\la 0\lp|\lp(a+\sqrt{\Nin}\rp)^{N+1}\lp(\ada+\sqrt{\Nin}\rp)^N\rp|0\rp\ra \\
&= \alpha\sum_{M=0}^N A_{N+1,M}\Nin^{2(N-M)},
\label{cpacs2}
\end{align}
where ${A_{N,M} = \frac{N!(N-1)!}{M!(N-M)!(N-M-1)!}\:(M=0,1,\cdots,N-1)}$.
The denominator is similarly calculated from ${\lp\la\sqrt{\Nin}\rp|a^N(\ada)^N\lp|\sqrt{\Nin}\rp\ra
= \sum_{M=0}^N B_{N,M}\Nin^{2(N-M)}}$, where ${B_{N,M} = \frac{\lp(N!\rp)^2}{M!\lp\{(N-M)!\rp\}^2}\:(M=0,1,\cdots,N)}$.
%
%

\begin{figure}
\includegraphics[clip,width=10cm]{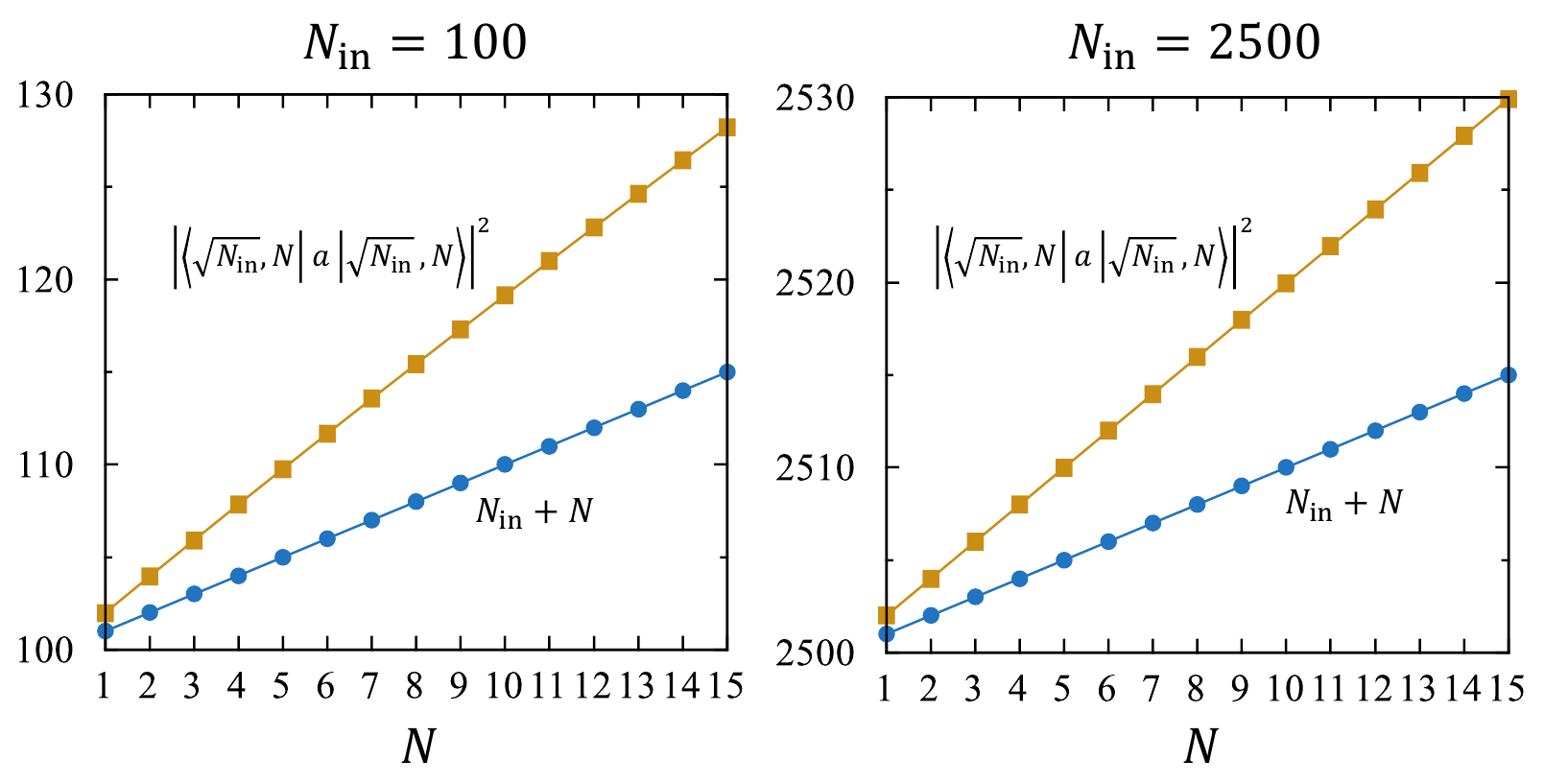}
\caption{Coherent photon numbers of the output pulse in our system with $N$ atoms (circles) and $N$-photon added state (squares) as a function of $N$.}
\label{fig:pacs}
\end{figure}

Figure\,\ref{fig:pacs} plots the coherent photon number in the $N$-PACS calculated from Eq.\,(\ref{cpacs}) with the coherent photon number in $|\sqrt{\Nin+N}\ra$, namely $\Nin+N$. It can be seen from the figure that the coherent photon number in the $N$-PACS is lager than $\Nin+N$ by $\sim N$, regardless of incident photon number. From this fact, it is obvious that the amplified (output) state in our study and $N$-PACS are in different states.
%
%
\section{Time evolution of correlation functions}
\label{sup:corfun}
\subsection{1-point function}
\label{1po}
To simulate the dynamics of the collective atomic spin, we introduce the Hilbert space spanned by ${|m\ra \equiv |J = \frac{N}{2}, J_z = m - \frac{N}{2}\ra}$, where $m$ ($= 0,1,\cdots, N$) denotes the number of excited atoms. This Hilbert space is enough to represent the wQED system considered in the paper, because the total angular momentum of the collective atomic spin is conserved. In this Hilbert space, the operators are expanded as
%
\begin{align}
\Jm &= \sum_{m=1}^N \sqrt{m(N-m+1)}\tau_{m-1,m}   ,
\label{jm}
\\
\Jp\Jm &= \sum_{m=1}^N m(N-m+1) \tau_{m,m}   ,
\label{jpjm}
\\
\Jz &= \sum_{m=0}^N (m-N/2)\tau_{m,m}   ,
\label{jz}
\end{align}
where $\tau_{m,m'}$ is defined as $\tau_{m,m'} \equiv |m\ra\la m'|$. Using the above equations, we derive the equations of motion for the 1-point function $\la\tau_{m,m'}\ra$: 
%
\begin{align}
\frac{d}{dt}s_{mm'} = \sum_{k,k'}D_{mm',kk'}s_{kk'}   ,
\label{eq:onep}
\end{align}
where $s_{mm'} \equiv \la\tau_{m,m'}\ra e^{i(m'-m)\omp t}$ and the input pulse is assumed as the form $\Ein(t)=\cE(t)e^{-i\omp t}$. $D_{mm',kk'}$ is non-zero only for several values of ${k,k'}$: $D_{mm',mm'} = {i(\oma-\omp)(m-m') - \frac{\gam m(N-m+1)}{2}-\frac{\gam m'(N-m'+1)}{2}}$, ${D_{mm',(m+1)(m'+1)} = \gam\sqrt{(m+1)(m'+1)(N-m)(N-m')}}$, ${D_{mm',(m-1)m'} = i\sqrt{\gam/2}\cE^*(t)\sqrt{m(N-m+1)}}$, ${D_{mm',m(m'+1)} = - i\sqrt{\gam/2}\cE^*(t) \sqrt{(m'+1)(N-m')}}$, ${D_{mm',(m+1)m'} = i\sqrt{\gam/2}\cE(t)\sqrt{(m+1)(N-m)}}$, and ${D_{mm',m(m'-1)} = - i\sqrt{\gam/2}\cE(t)\sqrt{m'(N-m'+1)}}$.
The expectation values of the system operators are calculated using $s_{mm'}$.
%
\subsection{2-point function and fluctuation in an output photon pulse}
\label{2po}
Each term in the quadrature fluctuations, $(\Delta\Xou)^2$ and $(\Delta\You)^2$, is calculated by
%
\begin{align}
\la\couda,\cou\ra &= \iint dt_1dt_2\fou(t_1)\fou^*(t_2)\la \Jp(t_1), \Jm(t_2)\ra   ,
\label{cdc}
\\
\la\cou,\cou\ra &= \iint dt_1dt_2\fou^*(t_1)\fou^*(t_2)\la \Jm(t_1), \Jm(t_2)\ra   ,
\label{cc}
\\
\la\couda,\couda\ra &= \iint dt_1dt_2\fou(t_1)\fou(t_2)\la \Jp(t_1), \Jp(t_2)\ra   .
\label{cdcd}
\end{align}
Here, in deriving the above formula, we used the input-output relation (\ref{aout}).
From Eq.(\ref{jm}), we have
%
\begin{align}
\la\Jp(t_1),\Jm(t_2)\ra &= \sum_{m,l}\sqrt{ml(N-m+1)(N-l+1)}
\nonumber
\\
&\times\la\tau_{m,(m-1)}(t_1),\tau_{(l-1),l}(t_2)\ra   ,
\label{Jp1Jm2}
\\
\la\Jm(t_1),\Jm(t_2)\ra &= \sum_{m,l}\sqrt{ml(N-m+1)(N-l+1)}
\nonumber
\\
&\times\la\tau_{(m-1),m}(t_1),\tau_{(l-1),l}(t_2)\ra   ,
\label{Jm1Jm2}
\\
\la\Jp(t_1),\Jp(t_2)\ra &= \sum_{m,l}\sqrt{ml(N-m+1)(N-l+1)}
\nonumber
\\
&\times\la\tau_{m,(m-1)}(t_1),\tau_{l,(l-1)}(t_2)\ra   .
\label{Jp1Jp2}
\end{align}
$\la\tau_{m,m'}(t_1),\tau_{l,l'}(t_2)\ra$ in the above equations are calculated by the 1-point correlation function $\la\tau_{m,m'}\ra$ and the 2-point correlation function $\la\tau_{m,m'}(t_1)\tau_{l,l'}(t_2)\ra$, which evolves as
%
\begin{align}
\frac{d}{dt_2}s^{(2)}_{mm',ll'} = \sum_{k,k'}D_{mm',kk'}(t=t_2)s^{(2)}_{kk',ll'}   ,
\label{eq:twop}
\end{align}
where ${s^{(2)}_{mm',ll'}(t_2,t_1) \equiv \la\tau_{l,l'}(t_1)\tau_{m,m'}(t_2)\ra e^{i\omp[(l'-l)t_1+(m'-m)t_2]}}$. The initial condition at ${t_2=t_1}$ is
%
\begin{align}
s^{(2)}_{mm',ll'}(t_1,t_1) = \delta_{l'm}s_{lm'}(t_1)   .
\label{ini_twop}
\end{align}
%
\section{Insights from a single-mode approach}
\label{sup:single}
%
\subsection{From traveling to standing waves}
\label{ssec:standing}
It is possible to gain some insight into the most notable features of this amplification process from a single-mode treatment, along the lines of the one adopted for the Jaynes-Cummings model in Refs.\,\cite{banacloche1991} and\,\cite{banacloche1992}. Note first that the field in the waveguide can be written in terms of standing instead of traveling waves, with modes described by operators $d_k = (a_k+b_k)/\sqrt 2$ and $e_k = (a_k-b_k)/\sqrt 2$ (see, for example, Section III of \cite{konyk}).  An atom at $r=0$ is maximally coupled to the $d$ modes, as Eq.\,(\ref{ham}) of the paper shows, (since they all have a maximum at that location), and not coupled at all to the $e$ modes, so that part of the field will not evolve. Since the transformation between the two sets of modes is essentially that of a beamsplitter, which combines coherent states in the same way as they do classical fields, it follows that, if the $d$ field remains approximately in a coherent state after the interaction, its changes will just be passed to the transmitted field afterwards:
%
\begin{align}
|\alpha\ra_a|0\ra_b = |\alpha/\sqrt 2\ra_d|\alpha/\sqrt 2\ra_e
\rightarrow |\beta\ra_d|\alpha/\sqrt 2\ra_e
= |\beta/\sqrt 2+\alpha/ 2\ra_a|\beta/\sqrt 2-\alpha/ 2\ra_b   .
\label{sta}
\end{align}
Suppose the standing-wave $d$ field, with amplitude $\alpha/\sqrt 2$, has one photon added, as well as a small phase change $\phi$.  Its amplitude will then be $\beta = \sqrt{\alpha^2/2+1}\,e^{i\phi} \simeq {\alpha/\sqrt 2} + 1/\sqrt 2\alpha +i \alpha\phi/\sqrt 2$.  When this is rewritten in terms of the traveling modes $a$ and $b$, as in Eq.\,(\ref{sta}), the original mode $a$ has an amplitude $\alpha + 1/2\alpha + i\alpha\phi/2$.  This means that the added photon is (with high probability) in the output field in the forward direction, since $(\alpha + 1/2\alpha)^2 \simeq \alpha^2 + 1$, whereas the phase shift is only half of the one calculated for the standing-wave mode. 

In the following we adopt a single-mode approach to estimate these changes in the field phase and amplitude. Although our system does not exhibit a net phase change, there is, as will be shown below, a growth in the phase fluctuations; this can be calculated for the standing-wave mode and then divided by 2 to get the traveling-wave result.

We start with a single-atom model, which can be solved exactly, and then we look at a less accurate approximation for the multiatom system, which will nevertheless allow us to estimate the phase spread observed in Fig.\,\ref{fig:flu}(c). 
%
\subsection{Single atom, single-mode approach}
\label{ssec:single}
A single-atom, single-mode system is known as the Jaynes-Cummings model\,\cite{jaynes}, and it is analytically solvable. With the atom initially in the excited state $|e\ra$, the state of the system at the time $t$ is 
%
\begin{align}
|\psi(t)\ra = \sum_{n=0}^\infty C_{n} &\lp[\cos(g\sqrt{n+1} \,t)|e\ra |n\ra -i \sin(g\sqrt {n+1} \,t) |g\ra |n+1\ra \rp]   ,
\label{JCevo}
\end{align}
where the $C_n = e^{-|\alpha|^2/2} \alpha^n/\sqrt{n!}$ are coherent-state coefficients. The mean number of photons in the initial state is $\bar{n} = |\alpha|^2$.  In what follows we will take $\alpha = \sqrt{\bar{n}}$ real for simplicity.

To bring the atom to the ground state $|g\ra$, we need $g t\sqrt{n+1}= \pi/2$, which cannot actually be satisfied for all values of $n$. We will define $t_\pi = \pi/(2 g\sqrt{\bar{n}})$, where $2g\sqrt{\bar{n}}$ plays the role of the Rabi frequency $\Om$ introduced in the main text of the paper, and look at what the state of the field is, conditioned on the assumption that the atom does in fact decay, that is, the last term on Eq.\,(\ref{JCevo}):
%
\begin{align}
\nonumber
|\Phi(t)\ra &= -i\sum_{n=0}^\infty C_{n} \sin(g\sqrt{n+1}\,t)|n+1\ra \\
&\simeq -\frac{1}{2} e^{igt/(2\sqrt{\bar{n}})}\sum_{n=0}^\infty C_{n}e^{ig\sqrt{n}\,t} |n+1\ra
+ \frac{1}{2} e^{-igt/(2\sqrt{\bar{n}})}\sum_{n=0}^\infty C_{n}e^{-ig\sqrt{n}\,t} |n+1\ra   ,
\label{JCevo2}
\end{align}
where we have expanded $\sqrt{n+1} \simeq \sqrt{n} + 1/(2\sqrt{n})$ and then replaced $n\rightarrow\bar{n}$ in the second term, since we expect that to be quite small for all the values of $n$ that matter in the sums in (\ref{JCevo2}) (assuming $\bar{n}$ is large enough). It can be shown that, as $\bar{n}$ increases, the norm of the state $|\Phi(t)\ra$, which gives the decay probability, approaches 1 at $t=t_\pi$.

Consider next the first of the two field states on the right-hand side of Eq.\,(\ref{JCevo2}) (a similar treatment applies to the second one). We can write
%
\begin{align}
\sum_{n=0}^\infty C_{n}e^{ig\sqrt{n}\,t}|n+1\ra
= V^\dag\sum_{n=0}^\infty C_{n}e^{ig\sqrt{n}\,t}|n\ra   ,
\label{JCevo3}
\end{align}
where $V^\dag=\sum_{n=0}^\infty |n+1\ra\la n|$ is one of the Susskind-Glogower operators\,\cite{susskind} that act as displacement operators for the photon number. If a phase operator $\hat \phi$, canonically conjugate to $\hat{n}$, existed, then one could write $V= e^{i\hat{\phi}}$ ($V^\dag = e^{-i\hat{\phi}}$) as the operators to decrease (increase) the photon number by 1\,\cite{loudon}, in the same way as $e^{-i\hat{p} a/\hbar}$ displaces a state's position by an amount $a$. As we shall see in a moment, for a coherent state with a sufficiently large amplitude, the action of $V$ and $V^\dag$ is indeed very similar to that of an ``exponential of phase'' operator.

As for the sums $\sum_{n=0}^\infty C_{n}e^{\pm ig\sqrt{n}\,t}|n\ra$, it was shown in Ref.\,\cite{banacloche1991} that they are well approximated (up to terms of order $1/\bar{n}$) by coherent states $e^{\pm ig\sqrt{\bar{n}}\,t/2} |\alpha e^{\pm igt/(2\sqrt{\bar{n}})}\ra$. We can then interpret the result (\ref{JCevo2}) as a sum of two coherent states, each with one photon added with respect to the original state, and with phases growing in opposite directions at the slow rates $\pm g/(2\sqrt{\bar{n}})$. By the time $t_\pi$ this is a very small phase shift, $\pm \pi/(4\bar{n})$, but for a large number of atoms it may be an observable effect, as we shall show below.

We have still to consider what the effect of the operator $V^\dag$ is on a coherent state. Let $|\psi^\pri\ra = V^\dag|\sqrt{\bar{n}} e^{i\phi}\ra$.  Direct calculation quickly shows that $\la\psi^\pri| \ada a |\psi^\pri\ra = \bar{n} + 1$, whereas for $\la\psi^\pri| a |\psi^\pri\ra$ we find
%
\begin{align} 
\la\psi^\pri| a |\psi^\pri\ra = e^{i\phi} \sum_n |C_n|^2 \frac{1}{\sqrt{\bar{n}}}\sqrt{n(n+1)}   ,
\end{align}
where the coefficients $C_n$ are those corresponding to the initial coherent state with mean photon number $\bar{n}$.  Expanding the term $\sqrt{n(n+1)}$ around $n=\bar{n}$, and using the known results for the Poisson distribution's variance, we obtain 
%
\begin{align}
\la\psi^\pri| a |\psi^\pri\ra \simeq e^{i\phi} \lp(\sqrt{\bar{n}} + \frac{1}{2\sqrt{\bar{n}}} - \frac{1}{8{\bar{n}}^{3/2}}\rp) \simeq \sqrt{\bar{n} + 1}\,e^{i\phi}   .
\end{align}
Accordingly, we can write $V^\dag|\sqrt{\bar{n}} e^{i\phi}\ra$ as a coherent state with amplitude $\sqrt{\bar{n} + 1}\,e^{i\phi}$ to a very good approximation.  

Collecting all these results, we have at the time $t_\pi = \pi/(2 g\sqrt{\bar{n}})$
%
\begin{align}
|\Phi(t)\ra \simeq -\frac 1 2 e^{i\pi/4} |\sqrt{\bar{n}+1} \,e^{i\pi/4\bar{n}}\ra + \frac 1 2 e^{-i\pi/4} |\sqrt{\bar{n}+1}\, e^{-i\pi/4\bar{n}}\ra   .
\label{JCevo4}
\end{align}
Since the intrinsic phase uncertainty of a coherent state is $1/(2\sqrt{\bar{n}})$, the phase split between the two components seen in Eq.\,(\ref{JCevo4}) will not really be resolvable for large $\bar{n}$, and it is in fact a good approximation to treat the whole thing as a single coherent state with amplitude $\sqrt{\bar{n}+1}$ and zero phase.  Nevertheless, we have chosen to write the state in the form (\ref{JCevo4}) to show that a broadening of the uncertainty in the phase quadrature is already hinted at in the single-atom case.  The following subsection shows that this is expected to be more pronounced in the multiatom case. 
%
\subsection{Multiatom case: semiclassical treatment}
Consider a Hamiltonian of the form ($\hbar =1$)
%
\begin{align}
H=g(a J_+ + J_- \ada)   ,
\label{appham}
\end{align}
where the $J_\pm$ operators are, as in the paper, angular momentum operators describing a collection of $N$ two-level atoms, with an equivalent total angular momentum $J=N/2$. If one replaces the field operators $a$ and $\ada$ in Eq.\,(\ref{appham}) by real constants, this Hamiltonian has as its stationary states the eigenstates of $J_x = \frac{1}{2}(J_+ +J_-)$, which we can write as $|m\ra_x$.  Accordingly, we may expect that if the initial field state is a coherent state with a large number of photons, the evolution of the fully quantum system will still be ``quasiclassical'' in some sense. Indeed, it was shown in Ref.\,\cite{banacloche1991}, for the case of a single atom, that the initial condition $|\Psi\ra = |\pm 1/2\ra_x|\alpha\ra$ led to approximately disentangled evolution, at the slow rate $g/2\sqrt{\bar{n}}$, where the field remained approximately coherent for short times, only with a phase $\phi = \mp gt/2\sqrt{\bar{n}}$. When starting from a state that is a superposition of the $\{|m\ra_x\}$ with equal weights for positive and negative $m$, we expect the field to end up in a superposition of positive and negative phases, explaining the phase spread that we see in Fig.\,\ref{fig:flu}(c).


This can be made more quantitative by extending slightly the analysis in the Appendix of Ref.\,\cite{banacloche1992}, to deal with the multiatom case.
Assuming an approximately factorizable field-atom evolution, the Heisenberg equations of motion for the Hamiltonian (\ref{appham}) can be written as
%
\begin{align}
\frac{d\la a\ra}{dt} &= -i g \la J_-\ra   , \label{app:em1} \\
\frac{d\la J_-\ra}{dt} &= 2ig \la J_z\ra\la a\ra   ,  \label{app:em2} \\
\frac{d\la J_z\ra}{dt} &= -i g\lp(\la J_+\ra\la a\ra -\la\ada\ra\la J_-\ra\rp)   . \label{app:em3}
\end{align}
Noting that $\la J_+\ra=\la J_-\ra^{*}$ and $\la\ada\ra=\la a\ra^{*}$, it is easy to see the system (\ref{app:em1})-(\ref{app:em3}) has two constants of the motion: $s=\sqrt{|\la J_-\ra|^2 + \la J_z\ra^2}$, and $N_{\rm tot} = |\la a\ra|^2 + \la J_z\ra$.  This suggests the following change of variables: let 
%
\begin{align}
\la a\ra &= r e^{i\phi}   , \label{app:a} \\
\la J_-\ra &= s \cos\eta \,e^{i\zeta}   , \label{app:Jm} \\
\la J_z\ra &= s \sin\eta   . \label{app:Jz}
\end{align}
where $r$ and $\eta$ are constrained by $r^2+s \sin\eta = N_{\rm tot}$.  This yields the system
%
\begin{align}
\dot\phi &=-\frac{gs}{r}\cos\eta\cos(\phi-\zeta)   , \label{app:ph} \\
\dot\zeta &= 2gr\tan\eta\cos(\phi-\zeta)   , \label{app:ze} \\
\dot\eta &= 2gr\sin(\phi-\zeta)   . \label{app:th}
\end{align}
When starting from a state $|m\ra_x$, the initial condition will be $\eta=0$, $s=m$, and $\phi=\zeta=0$ if the field's initial phase is zero.  An approximate result for short times is obtained by linearizing the system (\ref{app:ph})-(\ref{app:th}) assuming $\eta$ and $\phi$ remain small, and expanding $r$ around $\eta=0$ accordingly.  One obtains then, keeping only leading terms in powers of $1/\sqrt{\bar{n}}$,
%
\begin{align}
\phi(t) &\simeq \zeta(t) \simeq -\frac{g m t}{\sqrt{\bar{n}}}   , \label{app:ph2} \\
\eta(t) &\simeq -\frac{m}{\bar{n}}\sin^2(g\sqrt{\bar{n}}\, t)   . \label{app:th2}
\end{align}
This shows that, as anticipated, the field remains approximately coherent, with a phase that grows in one direction or another, dependent on the sign of $m$, while the atomic state stays close to the $x$-$y$ plane (at least for large $\bar{n}$, meaning small $\eta$), where it rotates at a rate that mirrors the field's. Comparison with the single-atom case treated above, with $m=\pm 1/2$, shows that this approach correctly predicts the evolving phase of the field for a state $|\pm\ra = \frac{1}{\sqrt{2}}(|e\ra \pm |g\ra)$, and hence the phase split when the initial state is $|e\ra = \frac{1}{\sqrt 2}(|+\ra +| -\ra)$.  (Unfortunately, due to its semiclassical nature, this approach cannot be used to correctly predict the change in photon number.)

Returning to the $N$ atom case, assume that the initial state is one where all the atoms are excited, that is, the $|J\ra_z$ state (with $J=N/2$).  We can write this as a superposition of $|m\ra_x$ states, and for each of them assume a phase drift as given by Eqs.\,(\ref{app:ph2}) and (\ref{app:th2}).  The overall phase spread (variance) can then be calculated as 
%
\begin{align}
\nonumber
\Delta^2\phi &= \sum_{m=-J}^J |{}_x\la m|J\ra_z|^2 \lp(\frac{g m t}{\sqrt{\bar{n}}}\rp)^2 \\
\nonumber
&=\frac{g^2 t^2}{\bar{n}} \sum_{m=-J}^J {}_z\la J| J_x |m\ra_x {}_x\la m| J_x |J\ra_z \\
\nonumber
&=\frac{g^2 t^2}{\bar{n}} {}_z \la J| J_x^2 |J\ra_z \\
&=\frac{Ng^2 t^2}{4\bar{n}}   ,
\end{align}
since ${}_z\la J| J_x^2 |J\ra_z=J/2 = N/4$.  This shows an uncertainty in the phase quadrature that grows as $\Delta\phi = \sqrt{N}gt/2\sqrt{\bar{n}}$. For a $\pi$ pulse, where $t=t_\pi = \pi/(2g\sqrt{\bar{n}})$, this becomes $\Delta\phi = \sqrt{N}\pi/(4\bar{n})$, inversely proportional to $\bar{n}$, and proportional to the square root of the number of atoms. 

To use this result for a traveling-wave field, with $\Nin$ photons, we need to do two things: divide the phase shift by 2, as explained in Eq.\,(\ref{sta}), and also realize that the mean number of photons $\bar{n}$ in the standing wave is actually equal to $\Nin/2$ (which also follows from Eq.\,(\ref{sta})). These two corrections cancel each other out, so we can use directly $\Delta\phi_{\rm ampl} = \sqrt{N}\pi/(4 \Nin)$ as an estimate for the increase in the phase uncertainty of the traveling-wave state due to the amplification process.  We can expect this to result in an increase in the fluctuations in the azimuthal quadrature approximately equal to $\Delta Y_{\rm ampl} \simeq \Delta \phi_{\rm ampl} \la c_{\rm out}\ra = \Delta \phi_{\rm ampl}\sqrt{G\Nin}$.  This should be added to the intrinsic quadrature uncertainty of the initial coherent state ($\Delta Y_0 = 1/2$) as ``independent noise,'' that is, adding the squares first and then taking the square root.

Altogether, then, the single-mode approach (with all the additional approximations, in particular the large $\bar{n}$ limit that leads to the approximate factorization of Eqs.\,(\ref{app:em1})-(\ref{app:em3}) ) predicts fluctuations in the azimuthal quadrature $\Delta Y$ after amplification given approximately by
%
\begin{align}
\Delta Y = \sqrt{\frac 1 4 + \frac{N \pi^2 G}{16 \Nin}}   .
\label{app:del_y}
\end{align}
The number of photons used in Fig.\,\ref{fig:flu}(c) can be inferred from the $\pi$ pulse condition, so for the rectangular pulse $\Nin= \pi^2/(2\gam \tp)$. At the three points where gain is highest, $\gam \tp = 0.1, 0.3,$ and $0.5$, we then get $\Delta Y = 0.63, 0.86,$ and $1.0$.  The first two values seem to agree fairly well with Fig.\,\ref{fig:flu}(c); the last one is too low, but at this point one only has $\sim 9.9$ photons in the pulse, and many of the approximations we have made are expected to break down.  The approximation is also not very good for the sine pulse (for which $\Nin = \pi^4/(16\gam \tp)$), predicting actually slightly lower values than for the rectangular pulse; however, again, it is not to be expected that a single-mode approximation could accurately predict effects depending on the pulse shape.


\begin{thebibliography}{99}
\bibitem{kimble}
H. J. Kimble, Nature (London) {\bf 453}, 1023 (2008).
%
%
\bibitem{meter2016}
R. Van Meter and S. J. Devitt, Computer {\bf 49}, 31 (2016).
%
\bibitem{simon}
C. Simon, Nat. Photonics {\bf 11}, 678 (2017).
%
\bibitem{roy}
D. Roy, C. M. Wilson, and O. Firstenberg, Rev. Mod. Phys. {\bf 89}, 021001 (2017).
%
\bibitem{chang}
D. E. Chang, J. S. Douglas, A. Gonz\'{a}lez-Tudela, C.-L. Hung, and H. J. Kimble, Rev. Mod. Phys. {\bf 90}, 031002 (2018).
%
\bibitem{turschmann}
P. T\"{u}rschmann, H. L. Jeannic, S. F. Simonsen, H. R. Haakh, S. G\"{o}tzinger, V. Sandoghdar, P. Lodahl, and N. Rotenberg, Nanophotonics {\bf 8}, 1641 (2019).
%
\bibitem{sheremet}
A. S. Sheremet, M. I. Petrov, I. V. Iorsh, A. V. Poshakinskiy, and A. N. Poddubny, arXiv:
2103.06824 (2021).
%
\bibitem{astafiev}
O. Astafiev, A. M. Zagoskin, A. A. Abdumalikov Jr., Yu. A. Pashkin, T. Yamamoto, K. Inomata, Y. Nakamura, and J. S. Tsai, Science {\bf 327}, 840 (2010).
%
\bibitem{fujiwara}
M. Fujiwara, K. Toubaru, T. Noda, H.-Q. Zhao, and S. Takeuchi, Nano Lett. {\bf 11}, 4362 (2011).
%
\bibitem{hoi2011}
I.-C. Hoi, C. M. Wilson, G. Johansson, T. Palomaki, B. Peropadre, and P. Delsing, Phys. Rev. Lett. {\bf 107}, 073601 (2011).
%
\bibitem{loo}
A. F. van Loo, A. Fedorov, K. Lalumiere, B. C. Sanders, A. Blais, and A. Wallraff, Science {\bf 342}, 1494 (2013).
%
\bibitem{kannan}
B. Kannan, M. Ruckriegel, D. Campbell, A. F. Kockum, J. Braumüller, D. Kim, M. Kjaergaard, P. Krantz, A. Melville, B. M. Niedzielski, A. Veps\"{a}l\"{a}inen, R. Winik, J. Yoder, F. Nori, T. P. Orlando, S. Gustavsson, and W. D. Oliver, Nature (London) {\bf 583}, 775 (2020).
%
\bibitem{mirhosseini}
M. Mirhosseini, E. Kim, X. Zhang, A. Sipahigil, P. B. Dieterle, A. J. Keller, A. Asenjo-Garcia, D. E. Chang, and O. Painter, Nature (London) {\bf 569}, 692 (2019).
%
\bibitem{sipahigil}
A. Sipahigil, R. E. Evans, D. D. Sukachev, M. J. Burek, J. Borregaard, M. K. Bhaskar, C. T. Nguyen, J. L. Pacheco, H. A. Atikian, C. Meuwly, R. M. Camacho, F. Jelezko, E. Bielejec, H. Park, M. Lon\v{c}ar, and M. D. Lukin, Science {\bf 354}, 847 (2016).
%
\bibitem{corozo2016}
N. V. Corzo, B. Gouraud, A. Chandra, A. Goban, A. S. Sheremet, D. V. Kupriyanov, and J. Laurat, Phys. Rev. Lett. {\bf 117}, 133603 (2016).
%
\bibitem{corozo2019}
N. V. Corzo, J. Raskop, A. Chandra, A. S. Sheremet, B. Gouraud, and J. Laurat, Nature {\bf 566}, 359 (2019).
%
\bibitem{goban}
A. Goban, C.-L. Hung, J. Hood, S.-P. Yu, J. Muniz, O. Painter, and H. Kimble, Phys. Rev. Lett. {\bf 115}, 063601 (2015).
%
\bibitem{solano}
P. Solano, P. Barberis-Blostein, F. K. Fatemi, L. A. Orozco, and S. L. Rolston, Nat. Comm. {\bf 8}, 1857 (2017).
%
\bibitem{dicke}
R. H. Dicke, Phys. Rev. {\bf 93}, 99 (1954).
%
\bibitem{gould}
R. G. Gould, ``The LASER, Light Amplification by Stimulated Emission of
Radiation''. In Franken, P.A.; Sands R.H. (Eds.) (1959). The Ann Arbor Conference on Optical Pumping, the University of Michigan, 15 June through 18 June 1959. p. 128. OCLC
02460155.
%
\bibitem{cummings}
F. W. Cummings, Phys. Rev. {\bf 140}, A1051 (1965).
%
\bibitem{kogelnik}
H. Kogelnik and C. V. Shank, Appl. Phys. Lett. {\bf 18}, 152 (1971).
%
\bibitem{mollow}
B. F. Mollow, Phys. Rev. A {\bf 5}, 2217 (1972).
%
\bibitem{einstein}
A. Einstein, Phys. Z. {\bf 18}, 121 (1917).
%
\bibitem{sun}
F. W. Sun, B. H. Liu, Y. X. Gong, Y. F. Huang, Z. Y. Ou, and G. C. Guo, Phys. Rev. Lett. {\bf 99}, 043601 (2007).
%
\bibitem{rephaeli}
E. Rephaeli and S. Fan, Phys. Rev. Lett. {\bf 108}, 143602 (2012).
%
\bibitem{valente}
D. Valente, Y. Li, J. P. Poizat, J. M. Gerard, L. C. Kwek, M. F. Santos, and A. Auffeves, New J. Phys. {\bf 14}, 083029 (2012).
%
\bibitem{fischer}
K. A. Fischer, OSA Continuum {\bf 1}, 772 (2018).
%
\bibitem{hoi2012}
I.-C. Hoi, T. Palomaki, J. Lindkvist, G. Johansson, P. Delsing, and C. Wilson, Phys. Rev. Lett. {\bf 108}, 263601 (2012).
%
\bibitem{zavatta}
A. Zavatta, S. Viciani, and M. Bellini, Science {\bf 306}, 660 (2004).
%
\bibitem{monsel}
J. Monsel, M. Fellous-Asiani, B. Huard, and A. Auff\'{e}ves, Phys. Rev. Lett. {\bf 124}, 130601 (2020).
%
\bibitem{francis}
J. T. Francis and M. S. Tame, Phys. Rev. A {\bf 102}, 043709 (2020).
%
\bibitem{agarwal}
G. S. Agarwal and K. Tara, Phys, Rev. A {\bf 43}, 492 (1991).
%
\bibitem{cave}
C. M. Caves, Phys. Rev. D {\bf 26}, 1817 (1982).
%
%
%
\bibitem{sup1}
See Sec.\,\ref{sup:connum} in the Supplemental Materials for derivation of the conservation condition of the total photon number.
%
\bibitem{sup2}
See Sec.\,\ref{sup:pacs} in the Supplemental Materials for comparison of the coherent photon number in the output state in our system with that of $N$-PACS.
%
\bibitem{sup3}
See Sec.\,\ref{1po} in the Supplemental Materials for derivation of the equations of motion.
%
\bibitem{sup4}
See Sec.\,\ref{2po} in the Supplemental Materials for derivation of the correlation functions.
%
%
%
\bibitem{exp}
We note that although some features of our results can be qualitatively understood from the single-mode models of cavity QED, including the noise addition in the phase direction (see Sec.\,\ref{sup:single} in the Supplemental Materials), it is in the context of waveguide-QED where we expect them to be most useful and better investigated.
%
\bibitem{grimsmo}
A. L. Grimsmo, J. Combes, and B. Q. Baragiola, Phys. Rev. X {\bf 10}, 011058 (2020).
%
\bibitem{leghtas}
Z. Leghtas, G. Kirchmair, B. Vlastakis, R. J. Schoelkopf, M. H. Devoret, and M. Mirrahimi, Phys. Rev. Lett. {\bf 111}, 120501 (2013).
%
\bibitem{grimm}
A. Grimm, N. E. Frattini, S. Puri, S. O. Mundhada, S. Touzard, M. Mirrahimi, S. M. Girvin, S. Shankar, and M. H. Devoret, Nature (London) {\bf 584}, 205-209 (2020).
\end{thebibliography}

\begin{thebibliography}{99}
\bibitem{banacloche1991}
J. Gea-Banacloche, ``Atom- and field-state evolution in the Jaynes-Cummings model for large initial fields,'' Phys. Rev. A {\bf 44}, 5913 (1991). 
%
\bibitem{banacloche1992}
J. Gea-Banacloche, ``A new look at the Jaynes-Cummings model for large fields: Bloch-sphere evolution and detuning effects,'' Opt. Commun. {\bf 88}, 531 (1992).
%
\bibitem{konyk}
W. Konyk and J. Gea-Banacloche, ``Quantum multimode treatment of light scattering by an atom in a waveguide,'' Phys. Rev. A {\bf 93}, 063807 (2016).
%
\bibitem{jaynes}
E. T. Jaynes and F. W. Cummings, Proc. IEEE {\bf 51}, 89 (1963).
%
\bibitem{susskind} L. Susskind and J. Glogower, Physica {\bf 1}, 49 (1964).
%
\bibitem{loudon} R. Loudon, \emph{The Quantum Theory of light}, 2nd Ed. (Oxford, 1983),  Sect. 4.8.
\end{thebibliography}
\end{document}